\documentclass[a4paper]{aa}
\usepackage{graphicx,amsfonts,amssymb,txfonts,natbib,latexsym}
\newcommand{\plotwd}{8.2cm}

\titlerunning{SDSS LRG acoustic oscillations} 
\title{Acoustic oscillations in the SDSS Luminous Red Galaxy sample power spectrum}

\author{G. H\"utsi}

\institute{Max-Planck-Institut f\"ur Astrophysik, Karl-Schwarzschild-Str. 1,
86740 Garching bei M\"unchen, Germany}

\offprints{G.H\"utsi , \\ \email{gert@mpa-garching.mpg.de}}
\date{Received / Accepted}
\begin{document}
\abstract {We calculate the redshift-space power spectrum of the Sloan
Digital Sky Survey (SDSS) Luminous Red Galaxy (LRG) sample, finding
evidence for a full series of acoustic features down to the scales of
$\sim 0.25\,h\,\mathrm{Mpc}^{-1}$. This corresponds up to the 9. peak
in the CMB angular power spectrum. The acoustic scale derived, $(107.8
\pm 4.3)\,h^{-1}\,\mathrm{Mpc}$ ($k \lesssim
0.2\,h\,\mathrm{Mpc}^{-1}$), agrees very well with the ``concordance''
model prediction and also with the one determined via the analysis of
the spatial two-point correlation function by
\citet{astro-ph/0501171}. This is not only an independent confirmation
 of \citet{astro-ph/0501171} results made with different methods and
software but also, according to our knowledge, the first determination of the power spectrum of the
SDSS LRG sample. By calculating the two-point correlation function
using the smooth cubic spline model fitted to the observed bandpowers
and comparing with the results of the direct determination we demonstrate the consistency of our results.
\keywords{large-scale structure of Universe }}
\maketitle

\section{Introduction}
In the beginning of $1970$'s it was already realized that acoustic
waves in the tightly coupled baryon-photon fluid prior to the epoch of
recombination will lead to the characteristic maxima and minima in the
post-recombination matter power spectrum. The same mechanism is also
responsible for the prominent peak structure in the CMB angular power
spectrum
\citep{1970Ap&SS...7....3S,1970ApJ...162..815P,1978SvA....22..523D}.
The scale of these features reflects the size of the sound horizon,
which itself is fully determined given the physical densities
$\Omega_b h^2$ and $\Omega_m h^2$. The acoustic horizon can be
calibrated using the CMB data, thus turning it into a standard ruler
which can be used to carry out standard cosmological tests. For
example, if we are able to measure the redshift and angular intervals
corresponding to the physically known acoustic scale in the matter
power spectrum at a range of redshifts, we can immediately find
angular diameter distance $d_{\rm A}$ and Hubble parameter $H$ as a
function of redshift. Having good knowledge of these dependencies
allows us to put constraints on the properties of the dark energy. To
carry out this project one needs a tracer population of objects whose
clustering properties with respect to the underlying matter
distribution is reasonably well understood. There have been several
works discussing the usage of galaxies
\citep{2003ApJ...594..665B,2003PhRvD..68f3004H,2003PhRvD..68h3504L,2003ApJ...598..720S}
and clusters of galaxies
\citep{2003PhRvD..68f3004H,2004ApJ...613...41M,astro-ph/0505441} for
this purpose. What is most important is that already currently
existing galaxy redshift surveys have lead to the detection of
acoustic features in the spatial distribution of galaxies, this way
providing clearest support for the standard gravitational instability
picture of the cosmic structure formation. In the paper by
\citet{astro-ph/0501171} the detection of the acoustic ``bump'' in the
two-point redshift-space correlation function of the SDSS
\footnote{http://www.sdss.org/} LRG sample is announced. The discovery
of similar features in the power spectrum of 2dF
\footnote{http://www.mso.anu.edu.au/2dFGRS/} galaxies is presented in
\citet{astro-ph/0501174}. These results clearly demonstrate the great
promise of the future dedicated galaxy redshift surveys like
K.A.O.S.\footnote{http://www.noao.edu/kaos/} Similarly, useful
measurements of the acoustic scale can be hoped by the planned SZ
cluster surveys like the ones carried out by the PLANCK Surveyor
\footnote{http://astro.estec.esa.nl/Planck} spacecraft and SPT
\footnote{http://astro.uchicago.edu/spt} \citep{astro-ph/0505441} and also with a large future photometric redshift surveys \citep{astro-ph/0411713}. In this paper we calculate the redshift-space power spectrum of the SDSS LRG sample finding evidence for the acoustic oscillations down to the scales of $\sim 0.25\,h\,\mathrm{Mpc}^{-1}$, which effectively correspond up to the 9. peak in the CMB angular power spectrum. These scales in the CMB are very strongly damped due to the finite width of the last-scattering surface and also due to the Silk damping \citep{1968ApJ...151..459S}. Also, at those scales the secondary CMB anisotropies (mostly thermal Sunyaev-Zeldovich effect \citep{1972CoASP...4..173S,1980ARA&A..18..537S}) start to dominate over the primary signal. On the other hand, features in the matter power spectrum, although being small ($\sim 5\%$ fluctuations), are preseved by the linear evolution and so opening up the way to probe acoustic phenomena at scales smaller than the ones accessible for the CMB studies. 

The paper is structured as follows. In Sec. 2 we describe the dataset to be analysed, Sec. 3 presents the method of the power spectrum calculation. The main results of this work are given in Sec. 4 and finally we conclude with Sec. 5.    
\section{Data}
We analyze the publicly available data from the SDSS Data Release 3
\citep{2005AJ....129.1755A}. Specifically, we carry out our power
spectrum measurements using the subset of the SDSS spectroscopic
sample known as the Luminous Red Galaxy (LRG) sample. The LRG
selection algorithm \citep{2001AJ....122.2267E} selects $\sim 12$
galaxies per square degree meeting specific colour and magnitude
criteria \footnote{For the exact details of the selection criteria see
\citet{2001AJ....122.2267E}}. The resulting set of galaxies consists
mostly of an early types populating dense cluster environments and as
such are significantly biased (bias factor $b \sim 2$) with respect to the underlying matter distribution. The selection method is very effective producing a galaxy sample with a reasonably high density up to the redshift of $z \sim 0.5$.

\begin{figure}
\centering
\includegraphics[width=\plotwd]
{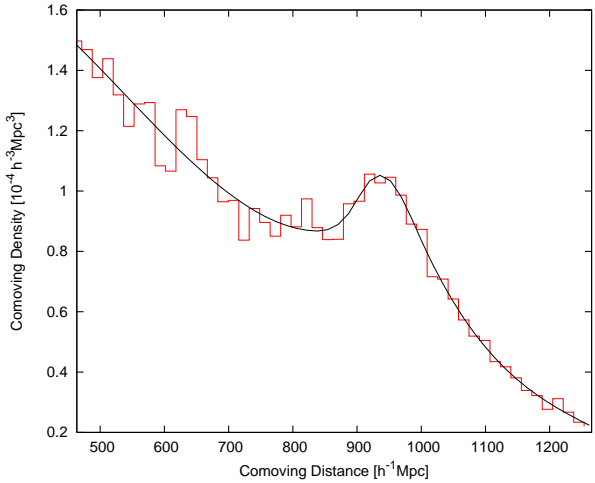}
\caption{Comoving number density of galaxies as a function of comoving distance. Smooth solid line shows a cubic spline fit to the number density estimated for 50 discrete radial bins.}
\label{fig1}
\end{figure}

\begin{figure}
\centering
\includegraphics[width=\plotwd]
{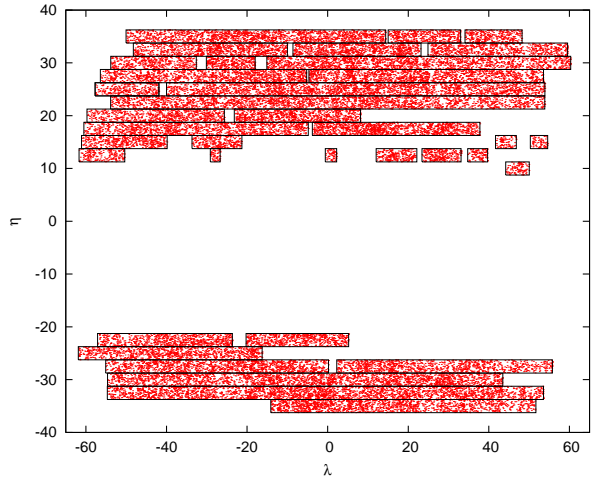}
\caption{Angular distribution of galaxies given in the SDSS survey coordinates $(\lambda,\eta)$. The survey mask is shown with solid lines.}
\label{fig2}
\end{figure}

Since the selection criteria are very complicated, involving both cuts
in magnitude and in color, and also due to the steepness of the
luminosity function the usual method of using only the luminosity
function to determine radial selection function does not work here
\citep{2005ApJ...621...22Z}. Here we simply build the radial selection
function as a smooth spline fit to the number density profile
as shown in Fig. \ref{fig1}. To calculate distances we choose the
cosmological parameters as given by the WMAP
\footnote{http://lambda.gsfc.nasa.gov/product/map/} 'concordance'
model \citep{2003ApJS..148..175S}. Unfortunately we could not find the
data describing properly the survey geometry \footnote{The data
related to the geometry seemed to be newer than the actual galaxy
data.} and so were forced to build the angular survey masks using the
galaxy data itself. As the number density of galaxies in the sample is
rather high, one can determine relatively accurately the beginning,
ending and also possible gaps in the scan stripes. The angular
distribution of the galaxies and also the boundaries of the survey
mask built in the above mentioned way is shown in
Fig. \ref{fig2}. Here the angular positions are plotted using the
so-called survey coordinate system of the SDSS \footnote{The
transformations between various coordinate systems used by the SDSS
are given e.g. in \citet{2002AJ....123..485S}.}. We apply lower and
upper redshift cutoffs of $0.16$ and $0.47$ as also done in
\citet{astro-ph/0501171}. In the analysis presented in this paper we
have excluded the three southern stripes since these just make the
survey window more anisotropic. We have also carried out calculations by including these stripes, finding out that the amount of additional information they provide is rather minor. Due to the very small sky coverage we have not used the data in slices 15, 42 and 43. So in total the analyzed galaxy sample covers $0.57 \,h^{-3}\,\mathrm{Gpc}^3$ over $2920$ square degrees on the sky and contains $37,998$ galaxies. 

\section{Power spectrum calculation}
\begin{figure}
\centering
\includegraphics[width=\plotwd]
{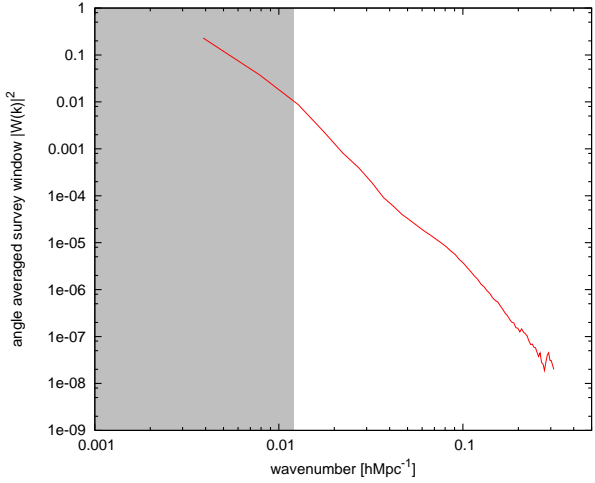}
\caption{Angle-averaged survey window $|W_k|^2$. The gray shaded area shows the region where the window is larger than $1\%$ of its maximum value.}
\label{fig3}
\end{figure}

We calculate the power spectrum using a direct Fourier method as
described in \citet{1994ApJ...426...23F} (FKP). Strictly speaking,
power spectra determined this way are the so-called pseudo spectra,
meaning that the estimates derived are convolved with a survey
window. Since in the case of the analysed LRG sample the volume
covered is very large, reaching $0.57 \,h^{-3}\,\mathrm{Gpc}^3$,  and
also the survey volume has relatively large dimensions along all
perpendicular directions, the correlations in the Fourier space are
rather compact. In general FKP estimator is working fine on
intermediate scales and in the case the power spectrum binning is wide
enough. More precisely, one can say that this estimator is fine for
wavenumbers $k$ and power spectrum bins $\Delta k$ larger than the
width of the survey window function $|W_{\mathbf{k}}|^2$, where
$W_{\mathbf{k}}$ is the Fourier transform of the product of the survey
mask and radial selection function. The angle average of this quantity
is shown in Fig. \ref{fig3}, where we have also marked with a gray
shaded stripe the scale where it is larger than $1 \%$ of its maximum value. This scale should serve as a rather conservative estimate of the effective window width. In the following we make sure that the power spectrum bins are chosen to be large enough so that the above discussed criteria are satisfied. To speed up the calculations of the power spectrum we make use of the Fast Fourier Transforms (FFTs), meaning that we need to build density field on a grid first. We use $512^{3}$ grid with a Triangular Shaped Cloud (TSC) mass assignment scheme \citep{1988csup.book.....H}. Very briefly the steps involved in calculating the power spectrum are as follows:
\begin{enumerate}
\item determination of the survey selection function (including the survey geometry) and mean underlying number density $\bar{n}$,
\item calculation of the overdensity field on a grid using TSC mass assignment scheme,
\item Fourier transformation of the gridded density field,
\item calculation of the raw 3D power spectrum,
\item subtraction of the shot noise component from the raw spectrum,
\item isotropization of the shot noise corrected 3D spectrum i.e. averaging over $k$-space shells,
\item application of the normalization correction due to selection effects,
\item deconvolving the smearing effect of the TSC mass assignment.
\end{enumerate}

More detailed description of our computational method with the results of application to a multitude of test problems can be found in \citet{astro-ph/0505441}. The power spectrum errors are estimated using the simple ``mode counting'' result of FKP (see also \citet{1998ApJ...499..555T}):
\begin{displaymath}
\frac{\Delta P}{P} = \sqrt{\frac{2}{V_{\rm eff}V_{\rm k}}}, 
\end{displaymath}
where $V_{\rm k}=4\pi k^2\Delta k/(2\pi)^3$ is the volume of the $k-$space shell and $V_{\rm eff}$ is the effective volume given by:
\begin{displaymath}
V_{\rm eff} = \frac{\left[\int w^2(z)\frac{{\rm d}V_{\rm c}}{{\rm d}z}{\rm d}z\right]^2}{\int w^4(z)\left[1+ \frac{1}{\bar{n}(z)P}\right]^2\frac{{\rm d}V_{\rm c}}{{\rm d}z}{\rm d}z}.
\end{displaymath}
Here $\bar{n}(z)$ is the mean number density of objects at redshift $z$, ${\rm d}V_{\rm c}$ is a comoving volume element and the weight function:
\begin{displaymath}
w(z)\propto \left\{ 
\begin{array}{lll}
\rm{const} & \rm{\quad for\ volume\ weighting}\\
\bar{n}(z) & \rm{\quad for\ number\ weighting}\\
\frac{\bar{n}(z)}{1+\bar{n}(z)P} & \rm{\quad for\ an\ optimal\ FKP\ weighting.}
\end{array} \right.
\end{displaymath}
In the following we use the FKP weight function, although due to the rather high number density and clustering strength of the sample the pure volume weighting is giving almost identical results.

\begin{figure}
\centering
\includegraphics[width=\plotwd]
{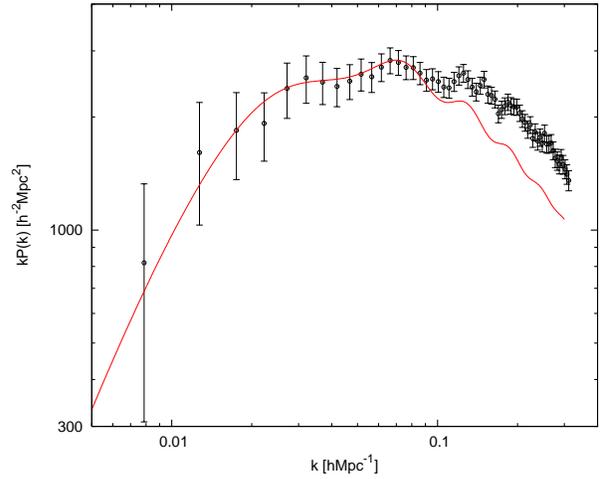}
\caption{Power spectrum of the SDSS LRG sample with the bin width $\Delta k \sim 0.005 \,h\,\mathrm{Mpc}^{-1}$. The solid line shows the linearly evolved matter power spectrum for the ``concordance'' cosmological model multiplied by the square of the bias parameter $b=1.95$.}
\label{fig4}
\end{figure}

\begin{figure}
\centering
\includegraphics[width=\plotwd]
{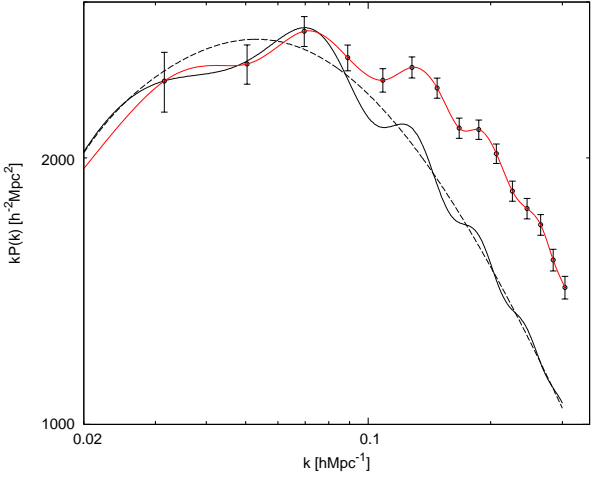}
\caption{Power spectrum of the SDSS LRG sample with the bin width $\Delta k \sim 0.02 \,h\,\mathrm{Mpc}^{-1}$. The upper solid line shows the smooth cubic spline model built through the data points, whereas the lower one corresponds to the same theoretical linearly evolved model as in Fig. \ref{fig4}. The dashed line displays the model with zero baryon content.}
\label{fig5}
\end{figure}

\section{Results}
\begin{figure}
\centering
\includegraphics[width=\plotwd]
{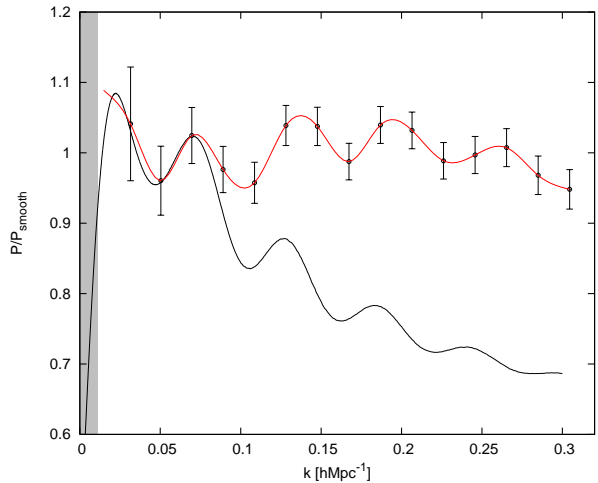}
\caption{The same spectra as in Fig. \ref{fig5}, except divided by the best fitting parabola in log-log coordinates, revealing clearly the oscillatory behaviour. The gray shaded stripe shows the effective width of the survey window as described in the main text.}
\label{fig6}
\end{figure}

In this section we present the main results of this work. In Fig. \ref{fig4} we show the redshift-space power spectrum of the above described SDSS LRG sample. We see that the power spectrum agrees remarkably well with the linearly evolved ``concordance'' model spectrum (shown with a solid line) down to the scales of $\sim 0.1 \,h\,\mathrm{Mpc}^{-1}$, beyond which the nonlinear evolution starts to boost the power. Here the bin width $\Delta k \sim 0.005 \,h\,\mathrm{Mpc}^{-1}$ is probably too narrow for the bins to be independent of each other. In Fig. \ref{fig5} we have increased the bin width up to $\sim 0.02 \,h\,\mathrm{Mpc}^{-1}$ and built a smooth cubic spline curve through the data points, revealing clear hints for the oscillatory behaviour of the spectrum. In addition we have plotted the same linearly evolved theoretical power spectrum as in the previous figure as well as the spectrum of a similar model but with a zero baryon content (shown by the dashed curve without any oscillatory behaviour). In order to display the detected fluctuations in a more clear manner we have fitted a smooth function in the form of a second order polynomial in log--log coordinates to the observed bandpowers. In Fig. \ref{fig6} we plot the same spectra as in the previous figure \footnote{With the only exception that we have not plotted a spectrum for a model with zero baryon content.} after dividing by this smooth function. The light gray shaded stripe shows the effective width of the survey window as described above. 

\begin{figure}
\centering
\includegraphics[width=\plotwd]
{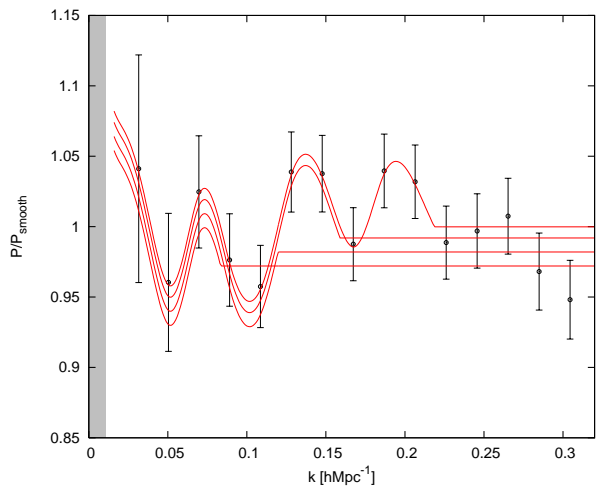}
\caption{Model spectra used to study the significance of the oscillatory features. Slight vertical shifts have been introduced in order to show clearly the locations of the transition scales applied.}
\label{fig7}
\end{figure}

\begin{figure}
\centering
\includegraphics[width=\plotwd]
{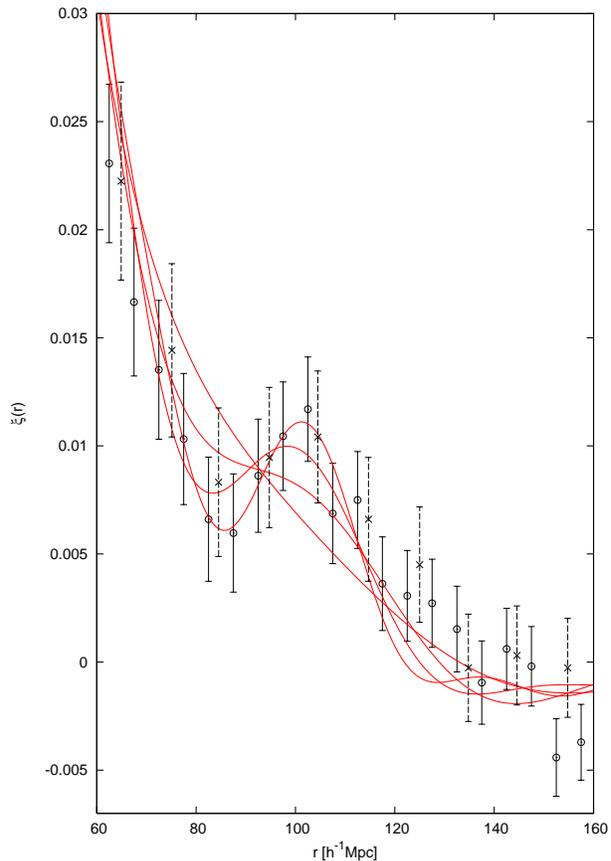}
\caption{Two-point correlation functions corresponding to the models shown in Fig. \ref{fig7}. The circles display the correlation function as determined directly from the data. The plotted errorbars are simply 2 times the Poissonian errors. And finally, the crosses with dashed errorbars show the correlation function found by \citet{astro-ph/0501171}.}
\label{fig8}
\end{figure}

It would be interesting to study how these oscillations in the power
spectrum translate to the peak in the two-point correlation function
discovered by \citet{astro-ph/0501171}. For this purpose we use the
cubic spline model described above and extend it outside of the
observed range by smoothly joining it to the power spectrum of the
``concordance'' model. The correlation function is now simply calculated as a Fourier transform of the power spectrum. To study the significance of the features at higher wavenumbers we have calculated correlation functions for several models that have oscillations ``switched off'' at various scales. The spectra of these models are shown in Fig. \ref{fig7}, where for the shake of clarity we have introduced slight vertical shifts between the curves so that the scales where the transition to the featureless spectrum takes place are easily visible. The corresponding correlation functions are given in Fig. \ref{fig8}. As expected, we see how the peak in the correlation function is getting broader and also decreasing in amplitude as we erase more features in the power spectrum. This clearly demonstrates the importance of many of the up-downs in the power spectrum to produce a relatively sharp feature in the two-point correlation function. Additionally, in this figure the circles with errorbars show the correlation function determined straight from the data using the edge-corrected estimator given by \citet{1993ApJ...412...64L}:
\begin{displaymath}
\xi (r) = \frac{DD - 2DR + RR}{RR},
\end{displaymath}
which has minimal variance for a Poisson process. Here DD, DR and RR represent the respective normalized data-data, data-random and random-random pair counts in a given distance range. Random catalogs were generated with 25 times the number of objects in the main catalog. The displayed errorbars are the simple Poissonian ones multiplied by the factor of 2. Since our aim here is not to perform any detailed correlation function study, this approximate treatment of errors, where  one simply multiplies the Poissonian errorbars (which underestimate true errors) by some ``fudge'' factor in the range $1.5-2.5$ (see e.g. \citet{1992ApJ...392..452M,1993A&A...280....5M}), should be well justified. The crosses with the dashed-line errorbars show the two-point function as determined by \citet{astro-ph/0501171}. We see that in general our results agree very well with their calculations.

\begin{figure}
\centering
\includegraphics[width=\plotwd]
{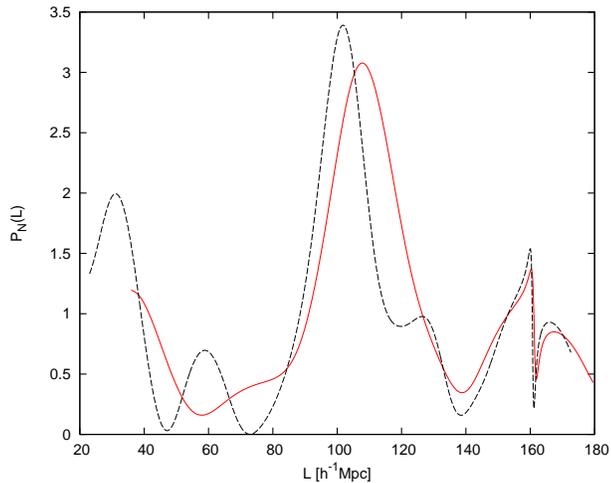}
\caption{The total variance normalized periodograms $P_N(L)=P_X(L)/\sigma ^2$ for (i) power spectrum points with wavenumbers up to $k \sim 0.2\,h\,\mathrm{Mpc}^{-1} $ (first 10 data points), (ii) all 15 data points. Case (i) is shown with a solid line.}
\label{fig9}
\end{figure}

To estimate the period of oscillations we calculate the periodogram $P_{X}(\omega)$ as described in \citet{1986ApJ...302..757H}:
\begin{displaymath}
P_{X}(\omega) = \frac{1}{2} \left \{ \frac{\left [ \sum \limits_{j=1}^{N_0}X(t_j)\cos \omega(t_j - \tau) \right ] ^2}{\sum \limits_{j=1}^{N_0}\cos ^2 \omega(t_j - \tau)} + \frac{\left [ \sum \limits_{j=1}^{N_0}X(t_j)\sin \omega(t_j - \tau) \right ] ^2}{\sum \limits_{j=1}^{N_0}\sin ^2 \omega(t_j - \tau)} \right \},
\end{displaymath}
where
\begin{displaymath}
\tan(2\omega \tau) = \frac{\sum \limits_{j=1}^{N_0}\sin 2 \omega t_j}{\sum \limits_{j=1}^{N_0} \cos 2 \omega t_j}
\end{displaymath}
and $X(t_j)$ represents a time series with $N_0$ data points. We calculate the frequency of oscillations for two different cases: (i) points with wavenumbers up to $k \sim 0.2\,h\,\mathrm{Mpc}^{-1} $ (first 10 data points), (ii) all 15 data points. The periodograms normalized by the total variance $P_N(\omega)=P_X(\omega)/\sigma ^2$ are shown in Fig. \ref{fig9}, where the solid line corresponds to the case (i). The uncertainty in the oscillation frequency is estimated as \citep{1986ApJ...302..757H}:
\begin{displaymath} 
\sigma_{\omega} = \frac{3\pi \sigma_{N}}{2\sqrt{N_0}TA}, 
\end{displaymath}
where $\sigma_N$ is the variance of the residuals after the signal has been subtracted, $T$ is the total length of the data set and $A$ is the amplitude of the signal. 

The acoustic scales measured in the above described manner are $(107.8 \pm 4.3)\,h^{-1}\,\mathrm{Mpc}$ and $(101.8 \pm 3.2)\,h^{-1}\,\mathrm{Mpc}$ for the cases (i) and (ii), respectively. These should be compared with the best-fit WMAP ``concordance'' model prediction of $106.5 \,h^{-1}\,\mathrm{Mpc}$. It is worth pointing out that the acoustic scale in the matter power spectrum is slightly larger than the corresponding scale in the CMB angular power spectrum due to the fact that acoustic waves do not stall completely at the epoch of decoupling, but keep on moving (although with a strongly reduced speed) up to the redshift $z \sim 100$ (the so-called drag-epoch).    

\section{Discussion and Conclusions}
In this paper we have calculated the redshift-space power spectrum of
the SDSS LRG sample, finding evidence for a series of acoustic
features down to the scales of $\sim
0.25\,h\,\mathrm{Mpc}^{-1}$. Using the obtained power spectrum we
predict the shape of the spatial two-point correlation function, which
agrees very well with the one obtained directly from the data. Also,
the directly calculated correlation function is consistent with the
results obtained by \citet{astro-ph/0501171}. We have made no attempts
to put constraints on the cosmological parameters, rather we have
assumed in our analysis the ``concordance'' cosmological model. The
derived acoustic scale $(107.8 \pm 4.3)\,h^{-1}\,\mathrm{Mpc}$ ($k
\lesssim 0.2\,h\,\mathrm{Mpc}^{-1}$) agrees well with the best-fit
WMAP ``concordance'' model prediction of $106.5
\,h^{-1}\,\mathrm{Mpc}$. Using all the data down to the wavenumbers $k
\sim 0.3\,h\,\mathrm{Mpc}^{-1}$, the obtained acoustic scale is
somewhat shorter, which might be due to the nonlinear effects
(transfer of power from large to small scales). 

The existence of the baryonic features in the galaxy power spectrum is
very important, allowing one (in principle) to obtain Hubble parameter
$H$ and angular diameter distance $d_A$ as a function of redshift,
this way opening up a possibility to constrain properties of the dark
energy \citep{2003PhRvD..68f3004H}. The currently existing biggest
redshift surveys, which are still quite shallow, do not yet provide
enough information to carry out this project fully. On the other hand,
it is extremely encouraging that even with the current generation of
redshift surveys we are already able to see the traces of acoustic
oscillations in the galaxy power spectrum, showing the great promise
for the dedicated future surveys like K.A.O.S. We have seen that
acoustic features seem to survive at mildly nonlinear scales ($k
\gtrsim 0.1\,h\,\mathrm{Mpc}^{-1}$), which is in agreement with the
results of the recent N-body simulations
\citep{2005Natur.435..629S,astro-ph/0507338}. In order to fully
exploit available information one needs a complete understanding of
how nonlinear effects influence these features. Nonlinear bias and
redshift space distortions also add extra complications. In general
redshift-space distortions, biasing and nonlinear evolution do not
create any oscillatory modulation in the power spectrum and so
acoustic features should be readily observable. On the other hand, only the effects that change the amplitude of the spectrum (e.g. redshift-space distortions on large scales) are easily factored out, but the ones involving stretching/compressing of the spatial scales do start to interfere with the cosmological distortions and so inhibit our ability to draw conclusions about the underlying cosmological model. So far there have been only a few works studying these important issues (e.g.  \citet{2005Natur.435..629S,astro-ph/0507338,astro-ph/0507307}) and probably it is fair to say that currently we really do not have a full theoretical description of them. This is also the reason why we have not tried to carry out any parameter estimation in this paper.   

The bare existence of the baryonic oscillations in the galaxy power spectrum tells us something important about the underlying cosmological model and the mechanism of the structure formation. First, it confirms the generic picture of the gravitational instability theory where the structure in the Universe is believed to be formed by the gravitational amplification of the small perturbations in the early Universe. Under the linear gravitational evolution all the density fluctuation modes evolve independently i.e. all the features in the power spectrum will be preserved. And certainly, we are able to identify features in the low redshift galaxy power spectrum that correspond to the fluctuations seen in the CMB angular power spectrum (which probes redshifts $z \sim 1100$), providing strong support for the above described standard picture of the structure formation. Actually, we can also probe scales that are unaccessible for the CMB studies due to the strong damping effects and steeply rising influence of the secondary anisotropies, reaching effectively the wavenumbers that correspond to the $8.-9.$ peaks in the CMB angular power spectrum. Second, the ability to observe baryonic features in the low redshift galaxy power spectrum demands rather high baryonic to total matter density ratio. In \citet{2003A&A...412...35B} it has been shown that it is possible to fit a large body of observational data with an Einstein--de Sitter type model if one adopts low value for the Hubble parameter and relaxes the usual assumptions about the single power law initial spectrum. In the light of the results obtained in our paper these models are certainly disfavored due to the fact that the high dark matter density completely damps the baryonic features. And finally, purely baryonic models are also ruled out since for them the expected acoustic scale would be roughly two times larger than observed here \footnote{For a clear discussion of this see Daniel Eisenstein's home page http://cmb.as.arizona.edu/$\sim$eisenste/acousticpeak/}. So the data seems to demand a weakly interacting nonrelativistic matter component and all the models that try to replace this dark matter component with something else e.g. modifying the laws of gravity might have hard time to fit these new observational constraints.  

\acknowledgements{I thank Rashid Sunyaev for valuable comments on the
    manuscript. I am very grateful to the support provided through the European Community's Human Potential Programme under contract HPRN-CT-2002-00124, CMBNET.\\
    Funding for the creation and distribution of the SDSS Archive has been provided by the Alfred P. Sloan Foundation, the Participating Institutions, the National Aeronautics and Space Administration, the National Science Foundation, the U.S. Department of Energy, the Japanese Monbukagakusho, and the Max Planck Society. The SDSS Web site is http://www.sdss.org/.

    The SDSS is managed by the Astrophysical Research Consortium (ARC) for the Participating Institutions. The Participating Institutions are The University of Chicago, Fermilab, the Institute for Advanced Study, the Japan Participation Group, The Johns Hopkins University, the Korean Scientist Group, Los Alamos National Laboratory, the Max-Planck-Institute for Astronomy (MPIA), the Max-Planck-Institute for Astrophysics (MPA), New Mexico State University, University of Pittsburgh, University of Portsmouth, Princeton University, the United States Naval Observatory, and the University of Washington.
}

\bibliographystyle{aa}
\bibliography{references}

\end{document}